\documentclass[preprint,showpacs,preprintnumbers,amsmath,amssymb]{revtex4}

\usepackage{graphicx}
\usepackage{dcolumn}
\usepackage{bm}

\begin{document}


\title{Superconductivity and superconducting order parameter phase fluctuations
in a weakly doped antiferromagnet}

\author{V.M.~Loktev}
\altaffiliation{Electronic address: vloktev@bitp.kiev.ua}
\affiliation{%
Bogolyubov Institute for Theoretical Physics,
Metrologicheskaya str. 14-b, Kiev-143, 03143 Ukraine}%

\author{V.M.~Turkowski}
\altaffiliation{Electronic address: turkowskiv@missouri.edu}
\affiliation{Department of Physics and Astronomy, University of
Missouri-Columbia, Columbia, Missouri 65211}%

\date{\today}

\begin{abstract}
The superconducting properties of a recently proposed
phenomenological model for a weakly doped antiferromagnet are
analyzed, taking into account fluctuations of the phase of the order
parameter. In this model, we assume that the doped charge carriers
can't move out of the antiferromagnetic sublattice they were
introduced. This case corresponds to the free carrier spectra with
the maximum at ${\bf k}=(\pm \pi /2 ,\pm \pi /2)$, as it was
observed in ARPES experiments in some of the cuprates in the
insulating state \cite{Damascelli}. The doping dependence of the
superconducting gap and the temperature-carrier density phase
diagram of the model are studied in the case of the
$d_{x^{2}-y^{2}}$ pairing symmetry and different values of the
effective coupling. A possible relevance of the results to the
experiments on high-temperature superconductors is discussed.
\end{abstract}



\maketitle

\section{Introduction}

Theoretical description of high-temperature superconductors (HTSCs)
remains one of the most important unsolved problems of modern
condensed matter theory. However, despite the lack of complete
understanding of this phenomena, some facts about the cuprate
materials are almost generally accepted by scientific community.
First of all, it is believed that superconductivity in this
materials is mainly generated in ${\rm CuO_{2}}$ layers, since even
samples with the layer spacing of $\sim 10\AA$ demonstrate very high
critical temperatures. Second, these materials in undoped regime are
antiferromagnetic insulators, which transform into superconductors
when the doping is larger than some critical value. Experiments
indicate the presence of strong antiferromagnetic correlations in
the superconducting regime \cite{Franck,Shen}. Third, the symmetry
of the superconducting order parameter is of a $d$-wave type in many
of the compounds \cite{Ding,Loeser}. It is believed by many
researches that these facts contain sufficient information to
understand the phenomena of HTSC. It is almost generally accepted
that this phenomena can be described in the framework of a simple
two-dimensional model of strongly correlated electrons (for a recent
over-review, see \cite{Lee}). The antiferromagnetic correlations
produce a BCS-type spin-wave pairing between electrons. One of the
most often studied model of HTSCs is the ${\rm t-J}$-model
\cite{Anderson,Zhang} (see, for example,
\cite{SchmittRink,Kane,Martinez,Plakida,Bensimon,Haule}). It was
shown that the solution of this model within different
approximations demonstrates $d$-wave superconductivity in the
underdoped regime. This and many other models, however, are rather
oversimplified, and they do not take into account some properties of
the cuprates. In particular, it is known from ARPES experiments that
the free carrier spectra of some cuprates like ${\rm
Sr_{2}CuO_{2}Cl_{2}}$ in the insulating phase has its maxima not at
the Brillouin zone edge at momenta ${\bf k}=(\pm \pi ,\pm \pi)$, but
at points ${\bf k}=(\pm \pi /2 ,\pm \pi /2)$ \cite{Damascelli}.

This fact suggests that the free carrier spectra must be
$\varepsilon ({\bf k})\sim -4t_{nn}\cos k_{x}\cos
k_{y}-2t_{nnn}(\cos 2k_{x}+\cos 2k_{y})$, i.e. the doped carriers
move only within one sublattice where they were born. This situation
is similar to the case of collinear antiferromagnetic dielectrics
with the Neel ground state, where electronic and spin excitations
can't move out of one sublattice to another. It is important that
the excitations must live on non-equivalent sublattices, i.e. on the
copper sublattices, which form an antiferromagnet. It is often
assumed for several reasons (see below), that doped holes in
cuprates occupy sites in the oxygen subsystem of the ${\rm CuO_{2}}$
planes \cite{Emery}. However, it is known that the n.n.n. hopping
parameter for this sublattice is much smaller than the n.n. hopping
parameter, which excludes the possibility to explain the spectra
experiments \cite{Damascelli} in this case. One can try to overcome
this difficulty by introducing the Zhang-Rice singlet, which
consists of a doped oxygen hole coupled to one of the closest copper
ions. The hole chooses the ion in such a way that the energy of the
singlet is minimal. Such states can move only within their magnetic
sublattices. However, these states are unstable due to several
reasons. First of all, the "up" and "down" spins of the copper ions
enter in the linear combination of the Zhang-Rice singlet in an
equivalent way, so a hole with the opposite spin projection will try
to form the same singlet with the neighboring localized ion spin
from another sublattice. Also, the oxygen ions occupy states which
are symmetrical with respect to the neighbor copper ions. Therefore,
the total exchange field which acts on a hole on the oxygen site is
compensated, and the hole becomes frustrated with respect to the
choice of the axis of the spin quantization. The proper axis of
quantization can be established, but in this case the hole
hybridized with the copper states will move in the ${\rm CuO_{2}}$
plane not feeling the magnetic ordering. This also doesn't allow one
to explain the spectra experiments.

These difficulties can be avoided by assuming that the doped holes
sit on the copper sites \cite{Loktev}. In fact, in the undoped state
the ${\rm CuO_{2}}$ planes consist of ions ${\rm Cu^{2+}}$ and ${\rm
O^{2-}}$. The doping leads to the valence change of some ions, which
results in appearance of movable charges in the system. The
assumption that the holes occupy the copper ions corresponds to the
transitions ${\rm Cu^{2+}\rightarrow Cu^{3+}}$, which are different
from usually considered transitions ${\rm O^{2-}\rightarrow O^{-}}$
in the case of the holes which occupy the oxygen sites. The reason
for the last assumption follows from the fact that it is believed by
many researches that the configuration ${\rm Cu^{3+}O^{2-}}$ has
higher energy comparing to ${\rm Cu^{2+}O^{-}}$. However, the
configuration ${\rm Cu^{3+}O^{2-}}$ can be preferable if one takes
into account the Coulomb attraction ${\rm V_{C}}$ inside the
configurations. In fact, since ${\rm V_{C}^{Cu^{3+}O^{2-}}\simeq
3V_{C}^{Cu^{2+}O^{-}}\simeq 6V_{C}^{Cu^{+}O^{-}}}$, and every copper
ion is surrounded by four oxygen ions, at the same time as every
oxygen is surrounded by only two copper ions, one can find that the
corresponding Coulomb attraction energy difference for two
configurations is of order ${\rm 20V_{C}^{Cu^{+}O^{-}}}$. This
difference in the Coulomb interaction energies can provide stability
of the ${\rm Cu^{3+}O^{2-}}$ configuration comparing to other
configurations, i.e. a possibility for the doped holes to move
within the copper sublattices. This picture corresponds
qualitatively to the ${\rm t-J}$-model case. However, it was usually
assumed that the free carrier dispersion relation in this model
corresponds to the nearest neighbor hopping, which does not allow
one to obtain the experimental spectra with the minimum at ${\bf
k}=(\pm \pi /2 ,\pm \pi /2)$.

A simple phenomenological Hamiltonian, which corresponds to charge
carriers that move within fixed sublattices, was proposed in order
to describe superconductivity in cuprates at low carrier densities
in \cite{Loktev}. The effective attraction between the doped
electrons on different sublattices was assumed to be equal to the
antiferromagnetic coupling $J$ between nearest site spins. In fact,
the carrier doping in the antiferromagnet leads to a breakdown of
the antiferromagnetic coupling $J$ between nearest site spins. This
results in an increasing of the energy of the system. This energy
increasing is minimal when two empty sites are the nearest
neighbors, because the minimal number of the antiferromagnetic
spin-spin (exchange) interaction bonds is broken in this case. This
phenomenological attraction was introduced for the first time by
Trugman in Ref.\cite{Trugman}. As it was shown in
Ref.~\cite{Scalapino}, the order parameter, which corresponds to the
pairs produced by such an interaction, has a $d$-wave symmetry.

Below, the doping dependence of the $d$-wave superconducting gap in
the model at different values of coupling and $T=0$ is studied. In
addition, we analyze the temperature-doping phase diagram of the
model. It is shown that the width of the pseudogap temperature
region between the mean-field critical temperature $T_{c}^{MF}$ and
the critical temperature of the Berezinskii-Kosterlitz-Thouless
(BKT) transition $T_{c}(\equiv T_{BKT})$ strongly depends on the
carrier concentration. We make a qualitative comparison of the
results with the experimental results on some HTSCs.

\section{Model}

The Hamiltonian of non-interacting doped d-hole carriers in HTSCs
can be approximated by
\begin{eqnarray}
H_{d}=\varepsilon_{d}\sum_{{\bf n}}\sum_{\sigma_{\bf n}} d_{{\bf
n}\sigma_{\bf n}}^{\dagger}d_{{\bf n}\sigma_{\bf n}}
-\frac{1}{2}\sum_{{\bf n},{\bf m}}\sum_{\sigma_n,\sigma_m} t_{{\bf
n}{\bf m}}\langle \sigma_{{\bf n}}| \sigma_{{\bf m}}\rangle d_{{\bf
n}\sigma_n}^{\dagger}d_{{\bf m}\sigma_m}, \label{Hd}
\end{eqnarray}
where $d_{{\bf n}\sigma_n}^{\dagger} (d_{{\bf n}\sigma_n})$ is the
creation (annihilation) operator of the electron on site ${\bf n}$
with spin $\sigma_{\bf n}$, $\varepsilon_{d}$ is the electron
on-site energy, $t_{{\bf n}{\bf m}}$ is the hopping parameter and
$\langle \sigma_{{\bf n}}| \sigma_{{\bf m}}\rangle$ is spin-spin
correlation function calculated in the ion system of coordinates
(see \cite{Loktev}). This Hamiltonian can be transformed to the
following form in terms of the Hubbard operators in the laboratory
system of spin coordinates:
\begin{eqnarray}
H_{d}=H_{coh}+H_{int}^{(1)}+H_{int}^{(2)},
\label{Hamiltonian1}
\end{eqnarray}
where
\begin{eqnarray}
H_{coh}=(\varepsilon_{d}-\mu)\sum_{{\bf n}}X_{{\bf n}}^{2,2}
-\frac{1}{2}\sum_{{\bf n},{\bf m}}t_{{\bf n}{\bf m}} \cos\frac{{\bf
Q}_{AFM}({\bf n}-{\bf m})}{2} X_{{\bf n}}^{2,1/2}X_{{\bf m}}^{1/2,2}
\label{Hcoh}
\end{eqnarray}
is a part of the Hamiltonian which
describes motion of free holes
in an antiferromagnetically ordered medium,
${\bf Q}_{AFM}{\bf a}=(\pm \pi ,\pm \pi)$, ${\bf a}$ is the lattice
constant
of a square lattice,
\begin{eqnarray}
H_{int}^{(1)}=-\frac{1}{2}\sum_{{\bf n},{\bf m}}t_{{\bf n}{\bf m}}
\sin\frac{{\bf Q}_{AFM}({\bf n}-{\bf m})}{2} \left( X_{{\bf
n}}^{2,1/2}X_{{\bf m}}^{1/2,2}S_{{\bf m}}^{-} -X_{{\bf
n}}^{2,1/2}X_{{\bf m}}^{1/2,2}S_{{\bf n}}^{+} \right) \label{Hint1}
\end{eqnarray}
and
\begin{eqnarray}
H_{int}^{(2)}=-\frac{1}{2}\sum_{{\bf n},{\bf m}}t_{{\bf n}{\bf m}}
\cos\frac{{\bf Q}_{AFM}({\bf n}-{\bf m})}{2} X_{{\bf
n}}^{2,1/2}X_{{\bf m}}^{1/2,2} S_{{\bf n}}^{+}S_{{\bf m}}^{-}
\label{Hint2}
\end{eqnarray}
describe non-coherent inter-ion hole transitions with one and two
spin excitations, correspondingly. $S_{{\bf n}}^{+}$ and $S_{{\bf
m}}^{-}$ are spin creation and annihilation operators. It is
important that expressions (\ref{Hamiltonian1})-(\ref{Hint2}) are
written in terms of the Hubbard operators, which directly take into
account the antiferromagnetic ordering in the system. The ion spin
projections in both magnetic sublattices are equal to $1/2$ in the
ground state of the crystal (we use local systems of coordinates for
each sublattice).
  The spin conservation is also taken into account,
which results in the fact that the electrons can move only on the
magnetic sublattice on which they were born (for details, see
\cite{Loktev}).

As it was mentioned in the Introduction, a simple effective
attraction between the doped electrons on different sublattices can
be introduced \cite{Trugman}. The carrier doping in the
antiferromagnet leads to an increasing of the energy of the system,
since it breaks the antiferromagnetic coupling $J$ between nearest
site spins. The energy increasing is minimal when two doped
particles occupy the nearest neighbor sites, since the minimal
number of the antiferromagnetic spin-spin bonds is broken in this
case. Therefore, the doping leads to an effective attraction between
carriers on different sublattices:
\begin{eqnarray}
H_{attr}=-J\sum_{{\bf n},{\bf \rho}={\bf a},{\bf b}}
X_{{\bf n}}^{2,2}X_{{\bf n}+{\bf \rho}}^{2,2}.
\label{Hattr}
\end{eqnarray}
The total Hamiltonian of the system is $H=H_{d}+H_{attr}$. For
simplicity, we neglect terms (\ref{Hint1}) and (\ref{Hint2}), which
do not contribute to a significant increasing of the superconducting
critical temperature. The first term corresponds to a BCS-like
interaction, which is small at low carrier densities, or Fermi
momenta of doped holes $k_{F}$. It is proportional to the energy of
spin waves $\Omega_{AFM}(k_{F})$, which is much smaller than the
exchange energy $J$. The term Eq.~(\ref{Hint2}), which corresponds
to two-magnon attraction, also doesn't contribute to a significant
increasing of the critical temperature in the $d$-wave pairing
channel \cite{Babichenko}.

Therefore, a simplified version of the Hamiltonian for the doped the
antiferromagnet can be written as:
\begin{eqnarray}
H=(\varepsilon_{d}-\mu)\sum_{{\bf n}}X_{{\bf n}}^{2,2}
-\frac{1}{2}\sum_{{\bf n},{\bf m}}t_{{\bf n}{\bf m}} \cos\frac{{\bf
Q}_{AFM}({\bf n}-{\bf m})}{2} X_{{\bf n}}^{2,1/2}X_{{\bf m}}^{1/2,2}
-J\sum_{{\bf n},{\bf \rho}={\bf a},{\bf b}} X_{{\bf n}}^{2,2}X_{{\bf
n}+{\bf \rho}}^{2,2}. \label{Hamiltonian2}
\end{eqnarray}

In the case of an antiferromagnet on a square lattice with two
sublattices, the free particle energy spectrum, which corresponds to
the first two terms in Eq.~(\ref{Hamiltonian2}), is:
\begin{equation}
\varepsilon ({\bf k})=\varepsilon_{d}
-4t_{2}\cos k_{x}\cos k_{y}
-2t_{3}(\cos 2k_{x}+\cos 2k_{y})-\mu ,
\label{epsilon}
\end{equation}
where $\mu$ is the chemical potential, and $t_{2}$ and $t_{3}$ are
the next nearest and next next nearest neighbor hopping parameters,
correspondingly. We use the units where the lattice constant is
equal to one: $a=1$. In order to have $\varepsilon ({\bf k})=0$ at
${\bf k}=0$ one can choose $\varepsilon_{d}=4t_{2}+4t_{3}$. The
chemical potential is connected with the free (or doped) particle
number in the system by the following natural relation:
\begin{equation}
\delta =\sum_{{\bf n}}\langle X_{{\bf n}}^{2,2}\rangle ,
\label{particlenumber}
\end{equation}
where the sum goes over two sublattices. The Hamiltonian
Eq.~(\ref{Hamiltonian2}) has a simpler structure, comparing to the
t-J-model Hamiltonian, yet it can describe some of the main physical
properties of underdoped cuprates.

\section{Zero-temperature properties}

To study superconducting properties of the system described by the
Hamiltonian Eq.~(\ref{Hamiltonian2}), it is convenient to introduce
generalized Nambu-Hubbard hole operators:
\begin{equation}
\Psi_{{\bf n}}(t)=
\left(
\begin{array}{c}
X_{{\bf n}}^{2,1/2}(t)\\
X_{{\bf n}}^{1/2 ,2}(t)
\end{array}\right) , \ \ \ \
\Psi_{{\bf n}}^{\dagger}(t)=
\left(
X_{{\bf n}}^{1/2 ,2}(t),
X_{{\bf n}}^{2,1/2}(t)
\right) ,
\label{Nambu}
\end{equation}
where ${\bf n}$ are lattice sites and $t$ is time. In this case the
time ordered Green function ${\hat G}_{{\bf n}{\bf m}}(t,t')=
-i\langle T (\Psi_{{\bf n}}(t)\Psi_{{\bf m}}^{\dagger}(t')\rangle$
is
\begin{equation}
{\hat G}_{{\bf n}{\bf m}}(t,t')=
-i\left(
\begin{array}{c}
\langle T X_{{\bf n}}^{2,1/2}(t) X_{{\bf m}}^{1/2,2}(t')\rangle\\
\langle T X_{{\bf n}}^{1/2,2}(t) X_{{\bf m}}^{1/2,2}(t')
\rangle
\end{array}
\begin{array}{c}
\langle T X_{{\bf n}}^{2,1/2}(t) X_{{\bf m}}^{2,1/2}(t')\rangle\\
\langle T X_{{\bf n}}^{1/2,2}(t) X_{{\bf m}}^{2,1/2}(t')\rangle
\end{array}
\right) .
\label{Gdef}
\end{equation}

The Green function Eq.~(\ref{Gdef}) satisfies the following equation
of motion:
\begin{equation}
i\frac{\partial}{\partial t}
{\hat G}_{{\bf n}{\bf m}}(t,t')
=\delta (t-t')\delta_{{\bf n}{\bf m}}{\hat I}
+\langle T[\Psi_{{\bf n}}(t),H]\Psi_{{\bf m}}^{\dagger}(t')
\rangle ,
\label{G}
\end{equation}
where, as it was mentioned above, $H$ is defined by
(\ref{Hamiltonian2}), and
\begin{equation}
{\hat I}=
\left(
\begin{array}{c}
\langle X_{{\bf n}}^{1/2,1/2}(t)+X_{{\bf n}}^{2,2}(t)\rangle\\
0
\end{array}
\begin{array}{c}
0\\
\langle X_{{\bf n}}^{1/2,1/2}+X_{{\bf n}}^{2,2}\rangle
\end{array}
\right) .
\label{I}
\end{equation}
This equality can be derived by using the commutation relations for
the Hubbard operators.

In the generalized mean-field theory approximation, the last term in
Eq.~(\ref{G}) can be linearized in the following way (see, for
example Ref.~\cite{Mancini}):
\begin{equation}
\langle T[\Psi_{{\bf n}},H]\Psi_{{\bf m}}^{\dagger}\rangle (\omega
)\simeq \sum_{{\bf l}}{\hat E}_{{\bf n}{\bf l}}{\hat G}_{{\bf l}{\bf
m}}(\omega ), \label{linearization}
\end{equation}
where
\begin{equation}
{\hat E}_{{\bf n}{\bf m}}=\langle \{ [\Psi_{{\bf n}},H], \Psi_{{\bf
m}}^{\dagger}\}\rangle  \label{linearization2}
\end{equation}
is the energy matrix. The nonlinear (dynamical) corrections to the
self-energy in Eq.~(\ref{linearization}) can be taken into account
\cite{Plakida}. We assume that the generalized mean-field
approximation Eq.~(\ref{linearization}) is good enough in the case
of low carrier densities, when the free quasi-particle excitations
correspond to the field represented by the Hubbard operators
Eq.~(\ref{Nambu}).

The expression for the energy matrix (\ref{linearization2}) can be
found by solving the Heisenberg equations of motion for the
$X$-operators. In terms of the energy matrix
Eq.~(\ref{linearization2}), the Green function can be written as:
\begin{equation}
{\hat G}_{{\bf n}{\bf m}}(\omega ) =\frac{{\hat I}}{\omega
\delta_{{\bf n}{\bf m}}-{\hat E}_{{\bf n}{\bf m}}}. \label{G2}
\end{equation}
To find the explicit expression for the Green function
Eq.~(\ref{G2}), let us write down the equations of motion for the
Hubbard operators:
\begin{eqnarray}
i\hbar\frac{\partial}{\partial t}X_{{\bf n}}^{1/2,2}(t)
=&~&(\varepsilon_{d}-\mu )X_{{\bf n}}^{1/2,2}
\nonumber \\
&-&\frac{1}{2}\sum_{{\bf l}} t_{{\bf nl}}\cos\frac{Q_{AFM}({\bf
n}-{\bf l})}{2} (X_{{\bf n}}^{1/2,1/2}+X_{{\bf n}}^{2,2})X_{{\bf
l}}^{1/2,2} -2J\sum_{{\bf \rho}} X_{{\bf n}+{\bf \rho}}^{2,2}X_{{\bf
n}}^{1/2,2}, \nonumber \\
\label{H1}
\end{eqnarray}
\begin{eqnarray}
i\hbar\frac{\partial}{\partial t}X_{{\bf n}}^{2,1/2}(t)
=&-&(\varepsilon_{d}-\mu )X_{{\bf n}}^{2,1/2}
\nonumber \\
&+&\frac{1}{2}\sum_{{\bf l}} t_{{\bf ln}}\cos\frac{Q_{AFM}({\bf
l}-{\bf n})}{2} X_{{\bf l}}^{2,1/2} (X_{{\bf n}}^{1/2,1/2}+X_{{\bf
n}}^{2,2}) +2J\sum_{{\bf \rho}} X_{{\bf n}}^{2,1/2}X_{{\bf n}+{\bf
\rho}}^{2,2}.
\nonumber \\
\label{H11}
\end{eqnarray}

Substitution of the expressions Eqs.~(\ref{H1}) and (\ref{H11})
instead of the anti-commutators (i$dX/dt=[X,H]$) into
Eq.~(\ref{linearization2}) and evaluation of the anti-commutators
give the following expression for the energy matrix:
\begin{equation}
{\hat E}_{{\bf nm}}={\tilde E}_{{\bf nm}}{\hat \tau}_{z} +{\tilde
\Delta}_{{\bf nm}}{\hat \tau}_{x}, \label{E}
\end{equation}
where ${\hat \tau}_{x}$ and ${\hat \tau}_{z}$ are the Pauli matrices
and
\begin{eqnarray}
{\tilde E}_{{\bf nm}}= &-&\delta_{{\bf nm}}(\varepsilon_{d}-\mu )
\langle X_{{\bf n}}^{1/2,1/2}+X_{{\bf n}}^{2,2}\rangle +\frac{1}{2}
\delta_{{\bf nm}} \sum_{{\bf l}}t_{{\bf ln}} \cos\frac{{\bf
Q}_{AFM}({\bf l}-{\bf n})}{2}\langle X_{{\bf l}}^{2,1/2}X_{{\bf
n}}^{1/2,2}\rangle
\nonumber \\
&+&2\delta_{{\bf nm}}\sum_{{\bf l}}J_{{\bf nl}}\langle X_{{\bf
n}}^{1/2,1/2} X_{{\bf l}}^{2,2}\rangle +\frac{1}{2}t_{{\bf mn}}
\cos\frac{{\bf Q}_{AFM}({\bf m}-{\bf n})}{2}\langle X_{{\bf
m}}^{1/2,1/2}(X_{{\bf n}}^{1/2,1/2})+X_{{\bf n}}^{2,2}) \rangle ,
\nonumber \\
 \label{Et}
\end{eqnarray}
\begin{eqnarray}
{\tilde \Delta}_{{\bf nm}}= -\frac{1}{2}\delta_{{\bf nm}} \sum_{{\bf
l}}t_{{\bf ln}} \cos\frac{{\bf Q}_{AFM}({\bf l}-{\bf n})}{2} \langle
X_{{\bf l}}^{2,1/2}X_{{\bf n}}^{2,1/2}\rangle -2J_{{\bf nm}}\langle
X_{{\bf n}}^{2,1/2}X_{{\bf m}}^{2,1/2}\rangle . \label{Deltat}
\end{eqnarray}
are the renormalized energy and the superconducting gap matrices.
Despite their complicated formal structure, it is possible to show
that the terms in Eq.~(\ref{Et}) and (\ref{Deltat}) have a very
simple physical interpretation. In particular, the second and the
third terms in Eq.~(\ref{Et}), proportional to $\delta_{\bf }$, lead
to a renormalization of the chemical potential $\mu\rightarrow \mu
'=\mu +\delta \mu$. The average $\langle X_{{\bf
m}}^{1/2,1/2}(X_{{\bf n}}^{1/2,1/2})+X_{{\bf n}}^{2,2}) \rangle$
multiplied by the hopping operator $t_{{\bf mn}}$ in the last term
of Eq.~(\ref{Et}) leads to renormalization of the quasiparticle
bandwidth in the limit of low doping. However, one can put $\langle
X_{{\bf m}}^{1/2,1/2}(X_{{\bf n}}^{1/2,1/2})+X_{{\bf n}}^{2,2})
\rangle\simeq 1$ in this limit, since in the limit of low carrier
concentrations $X_{{\bf n}}^{1/2,1/2})+X_{{\bf n}}^{2,2}\simeq 1$,
and the renormalization of the quasiparticle band is not strong.
Therefore, the energy function Eq.~(\ref{Et}) can be approximated by
the free energy spectra expression Eq.~(\ref{epsilon}) multiplied by
$-1$ in the momentum space. It is also assumed that the chemical
potential in Eq.~(\ref{epsilon}) is renormalized.

The expression for the gap function Eq.~(\ref{Deltat}) can be also
simplified. In fact, as it was shown in Ref.~\cite{Scalapino}, the
attraction (\ref{Hattr}) favours a superconducting pairing with the
$d$-wave symmetry of the order parameter. Therefore, we assume that
the strongest pairings in the system takes place in the $d$-wave
pairing case, and neglect the first term in Eq.~(\ref{Deltat}),
which doesn't contribute to the pairing in the d-wave channel.
Therefore, the gap function can be approximated in the following way
in the momentum representation:
\begin{eqnarray}
\Delta ({\bf k})= -2\sum_{{\bf q}} J({\bf k}-{\bf q}) \langle
X_{-{\bf q}}^{2,1/2}X_{{\bf q}}^{2,1/2} \rangle , \label{Deltak}
\end{eqnarray}
where we introduced the nearest neighbor attraction kernel $ J({\bf
k})=2J\left[ \cos (k_{x}-q_{x})+\cos (k_{y}-q_{y})\right]$. Thus,
Green function (\ref{G2}) has the following form in the momentum
space:
\begin{equation}
G(\omega ,{\bf k})=\frac{1}{\omega +\varepsilon ({\bf k})\tau_{z}
-\Delta({\bf k})\tau_{x}}, \label{GF}
\end{equation}
where we assumed that in the limit of low carrier densities the
normalization matrix is approximately equal to the unit matrix
${\hat I}\simeq {\hat 1}$.

To find the unknown gap function $\Delta (\bf k)$ and the
renormalized chemical potential $\mu '$, one can write down and
solve the system of equations for these functions by using the
fluctuation-dissipation theorem:
\begin{equation}
\langle A B\rangle =\frac{1}{\pi}\int_{-\infty}^{\infty}
d\omega\frac{d\omega \Im G_{AB}(\omega )}{e^{(\omega -\mu )/T}+1}.
\label{fluctuationdissipation}
\end{equation}

The zero-temperature equations which connect $\Delta ({\bf k})$ and
$\mu$ with the parameters $t_{2}$, $t_{3}$, $J$ and $\delta$ follow
from Eqs.~(\ref{Deltak}) and (\ref{particlenumber}):
\begin{equation}
\Delta ({\bf k})=-2\sum_{\bf q} J({\bf k}-{\bf q}) \frac{\Delta
({\bf q})} {\sqrt{\varepsilon^{2}({\bf q})+\Delta^{2}({\bf q})}},
\label{gapequation}
\end{equation}
\begin{equation}
\delta =\sum_{{\bf k}}\left[1+\frac{\varepsilon ({\bf k})}
{\sqrt{\varepsilon^{2}({\bf k})+\Delta^{2}({\bf k})}}\right] .
\label{numberequation}
\end{equation}

Since we consider the case, when the pairing in the system takes
place in the $d$-wave channel, we put $\Delta ({\bf
k})=\Delta_{d}\gamma_{d}({\bf k})$, where $\gamma_{d}({\bf k})=(\cos
k_{x}-\cos k_{y})$, in Eqs.~(\ref{gapequation}) and
(\ref{numberequation}). In order to these equations in this case,
one must extract the $d$-wave piece form the interaction kernel and
approximate it by this function: $J({\bf k}-{\bf q})\rightarrow
2J\gamma_{d}({\bf k})\gamma_{d} ({\bf q})$. In this case,
Eq.~(\ref{gapequation}) for the superconducting order parameter
acquires a rather simple form
\begin{equation}
1=4J\sum_{\bf q}\gamma_{d}^{2}({\bf q}) \frac{1}
{\sqrt{\varepsilon^{2}({\bf q})+\Delta_{d}^{2}\gamma_{d}^{2}({\bf
q})}}. \label{gapequation2}
\end{equation}

The solution of the set of Eqs. ~(\ref{numberequation}) and
(\ref{gapequation2}) at different values of interaction and hopping
$t_{3}$ is presented in Fig.1. As it follows from this Figure, the
gap is not very sensitive to the values of the next-nearest neighbor
hopping parameter, but it strongly depends on the interaction
potential. Superconductivity is suppressed when the carrier density
is smaller than some critical value. This value is also increasing
when the effective attractive interaction $J$ is decreasing. This
situation is, in principle, similar to the case with attracting
electrons, when there is no antiferromagnetic background for the
carrier motion (see, for example \cite{IJMP,PhysicaC}).

\begin{figure}[h]
\centering{
\includegraphics[width=7.2cm]{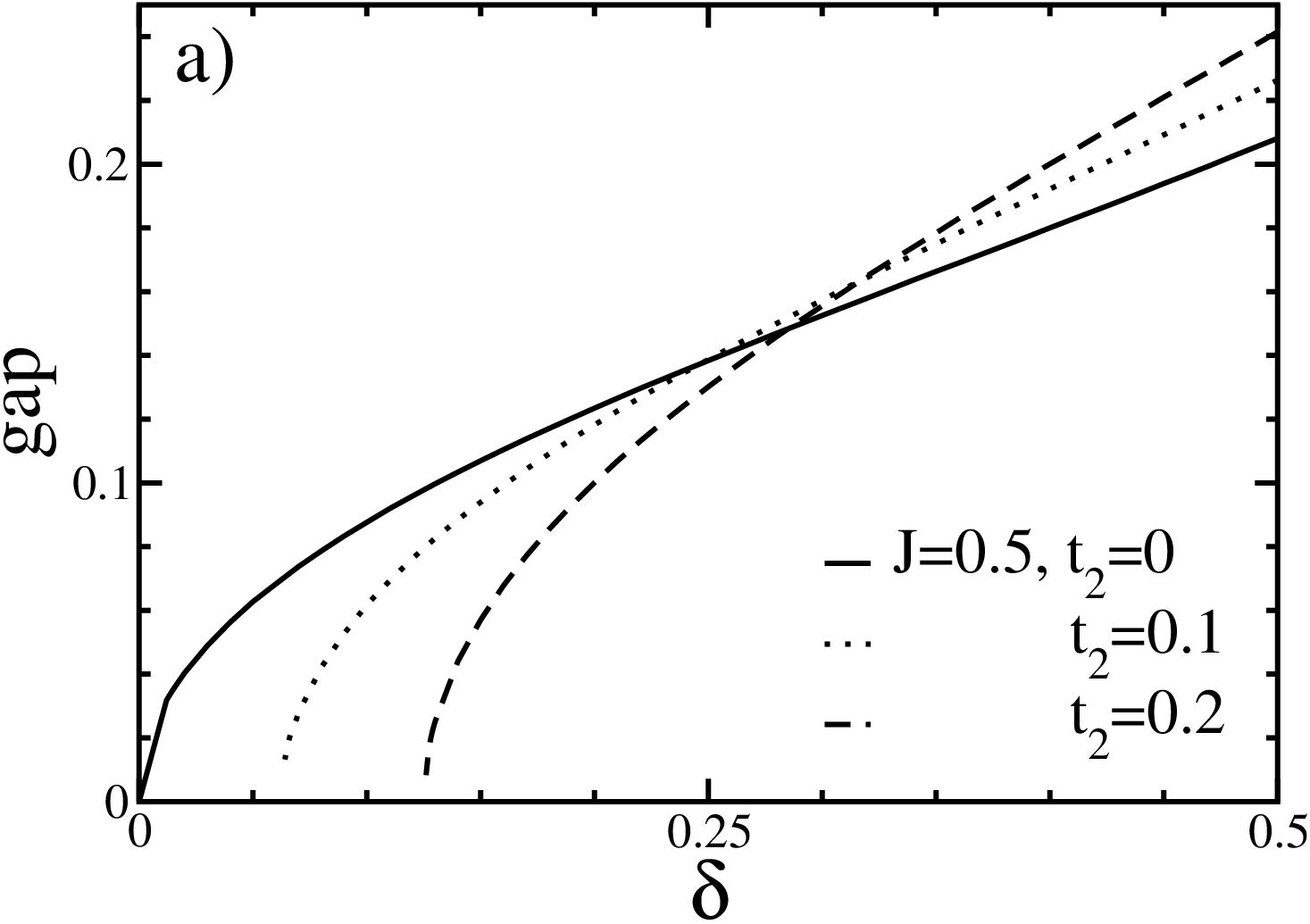}
.\\
.\\
\includegraphics[width=7.2cm]{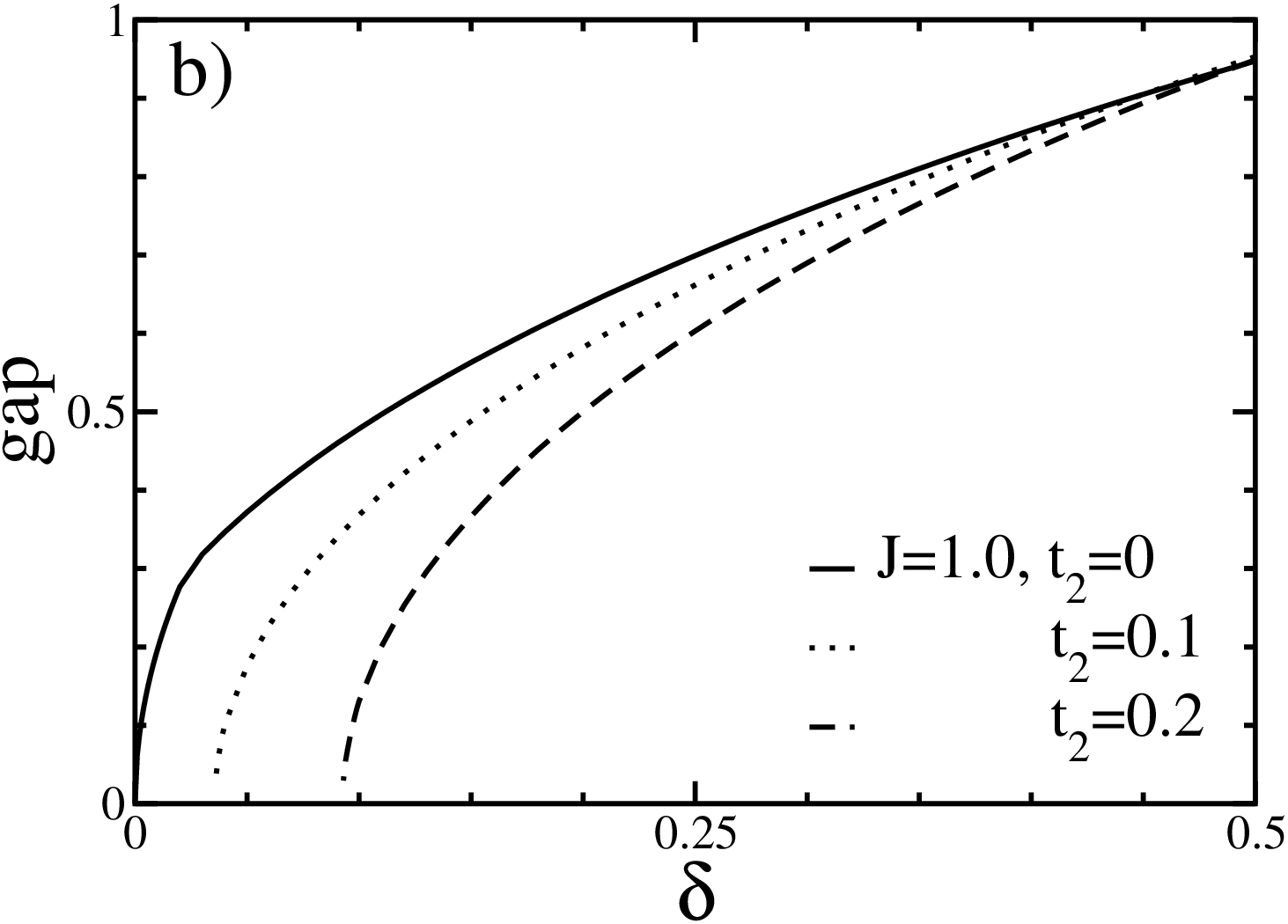}
} \caption{The gap as function of carrier density at $t_{3}=1$, and
different values of $t_{2}$ and $J$. Here and below all functions
are expressed in units of $t_{3}$. Here and in Fig.2b), we don't
present some results in the case when the gap or the critical
temperature is very small, since it was difficult to get accurate
numerical solutions in these cases.} \label{fig:1}
\end{figure}

\section{Temperature-carrier density phase diagram}

In this Section, we study the finite temperature properties of the
model. It is known that there are two critical temperatures in the
two-dimensional superconducting systems: the mean-field critical
temperature $T_{c}^{MF}$, below which the uncorrelated pairs start
to form, and the Berezinskii-Kosterlitz-Thouless critical
temperature $T_{BKT}<T_{c}^{MF}$, below which the phases of the pair
wave functions become algebraically ordered (for over-review, see
 for example \cite{PhysRep}).
Since such an order is the only possible order in our system, as it
was stated above, we have to put $T_{BKT}=T_{c}$.

\subsection{The mean-field critical temperature}

To find the dependence of the critical temperature $T_{c}^{MF}$ on
the particle density in the d-wave pairing channel, we need to solve
the finite-temperature version of the set of
Eqs.~(\ref{gapequation}) and (\ref{numberequation}). These equations
follow from (\ref{fluctuationdissipation}) , (\ref{particlenumber})
and (\ref{Deltak}):
\begin{equation}
1=4J\sum_{\bf q}\gamma_{d}^{2}({\bf q}) \tanh \left(
\frac{\sqrt{\varepsilon^{2}({\bf
q})+\Delta_{d}^{2}\gamma_{d}^{2}({\bf q})}}{2T}\right) \frac{1}
{\sqrt{\varepsilon^{2}({\bf q})+\Delta_{d}^{2}\gamma_{d}^{2}({\bf
q})}}, \label{gapequationT}
\end{equation}
\begin{equation}
\delta =\sum_{{\bf k}}\left[1+\tanh \left(
\frac{\sqrt{\varepsilon^{2}({\bf
q})+\Delta_{d}^{2}\gamma_{d}^{2}({\bf
q})}}{2T}\right)\frac{\varepsilon ({\bf k})}
{\sqrt{\varepsilon^{2}({\bf k})+\Delta_{d}^{2}\gamma_{d}^{2}({\bf
k})}}\right] . \label{numberequationT}
\end{equation}

The system of equations for $T_{c}^{MF}$ and $\mu '$ can be obtained
from Eqs.~(\ref{gapequationT}) and (\ref{numberequationT}) by
putting the amplitude of the order parameter equal to zero. The
solution of these equation shows that the doping-dependence of the
mean-field critical temperature (Figs.2-4) have qualitatively the
same form as the zero-temperature gap-dependence.

\begin{figure}[h]
\centering{
\includegraphics[width=6.0cm]{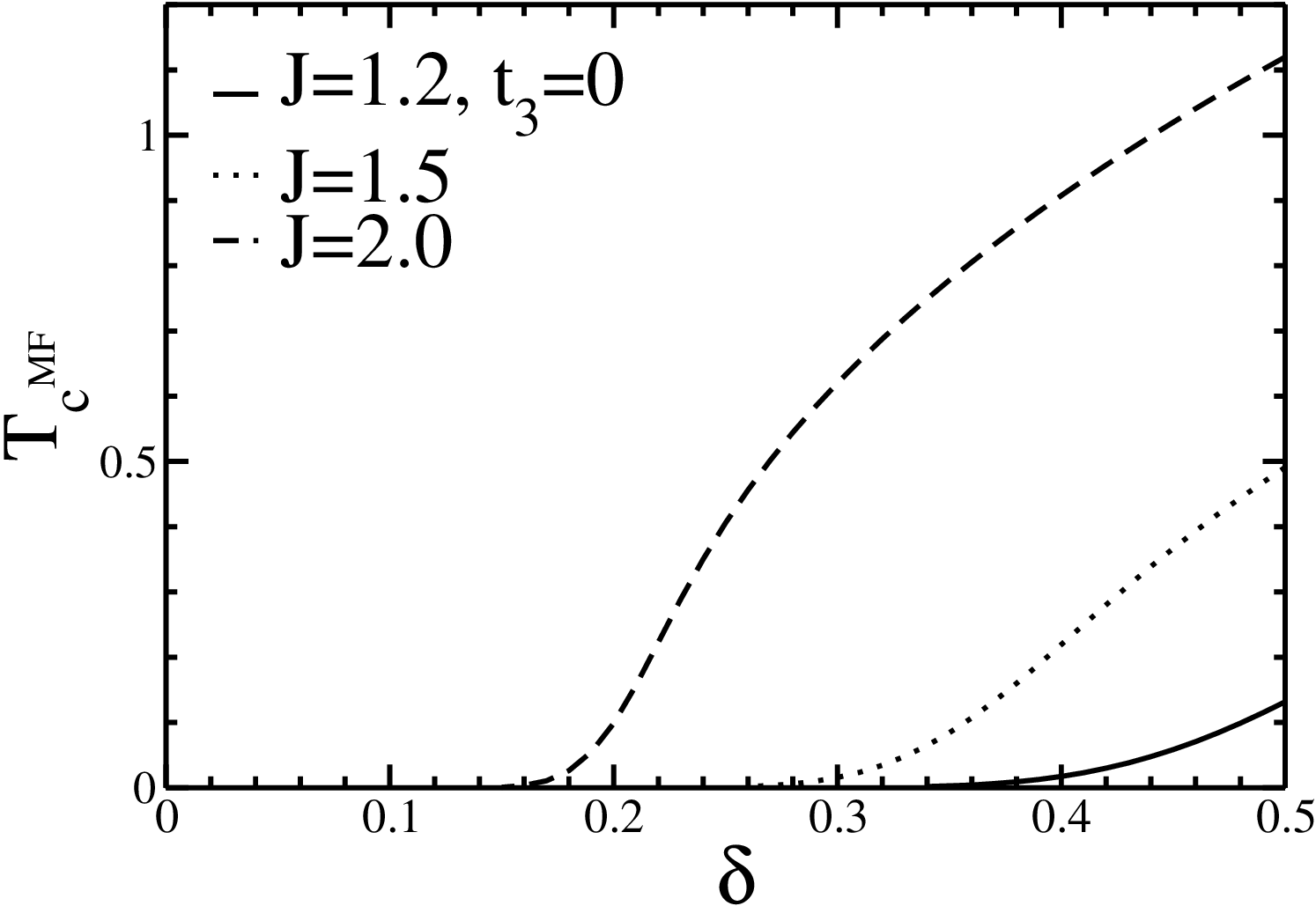}
\includegraphics[width=6.0cm]{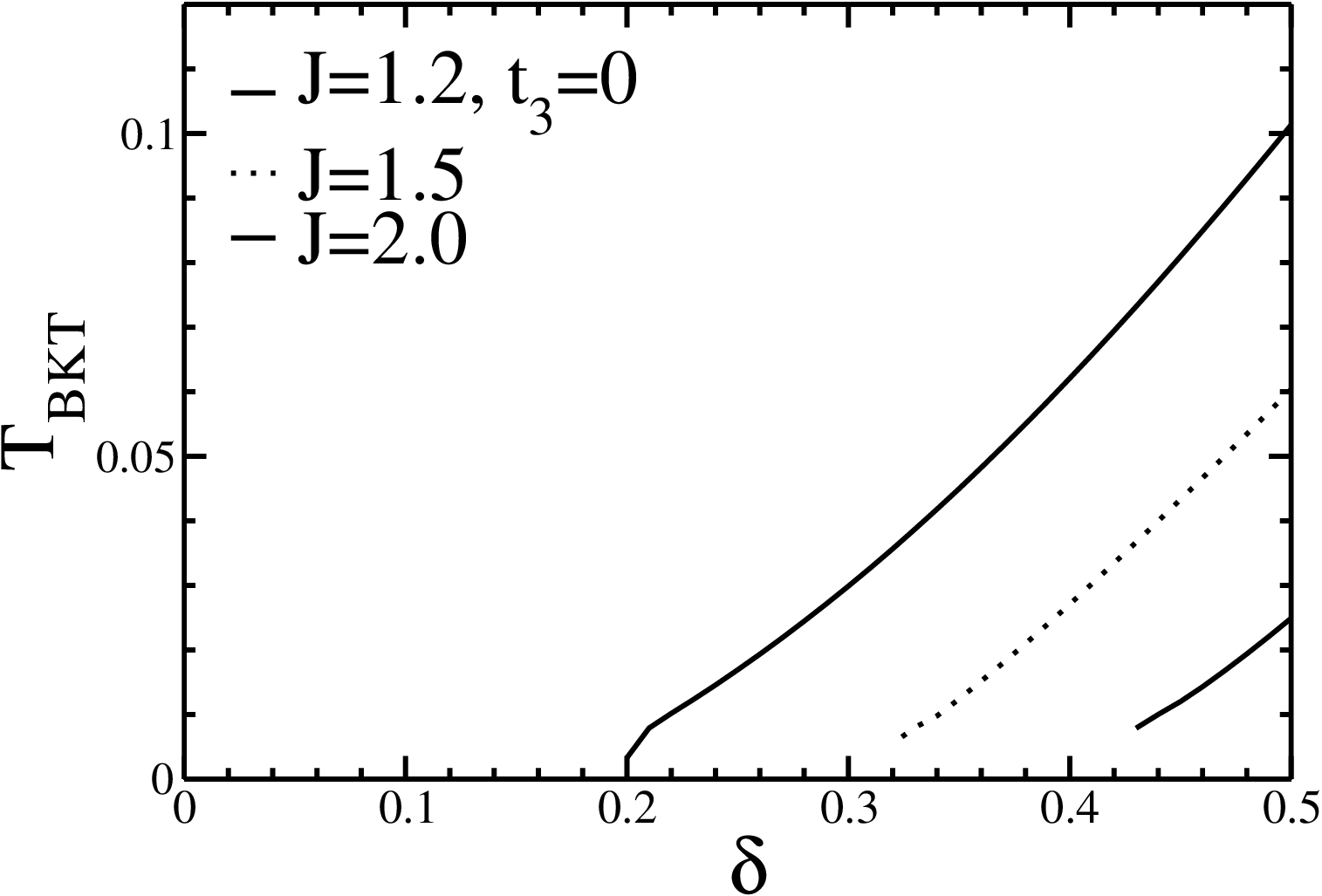}
} \caption{The mean-field (left) and the BKT (right) critical
temperatures as functions of carrier density at $t_{2}=1$, $t_{3}=0$
and different values of $J$.}
\label{fig:2}
\end{figure}

\begin{figure}[h]
\centering{
\includegraphics[width=6.0cm]{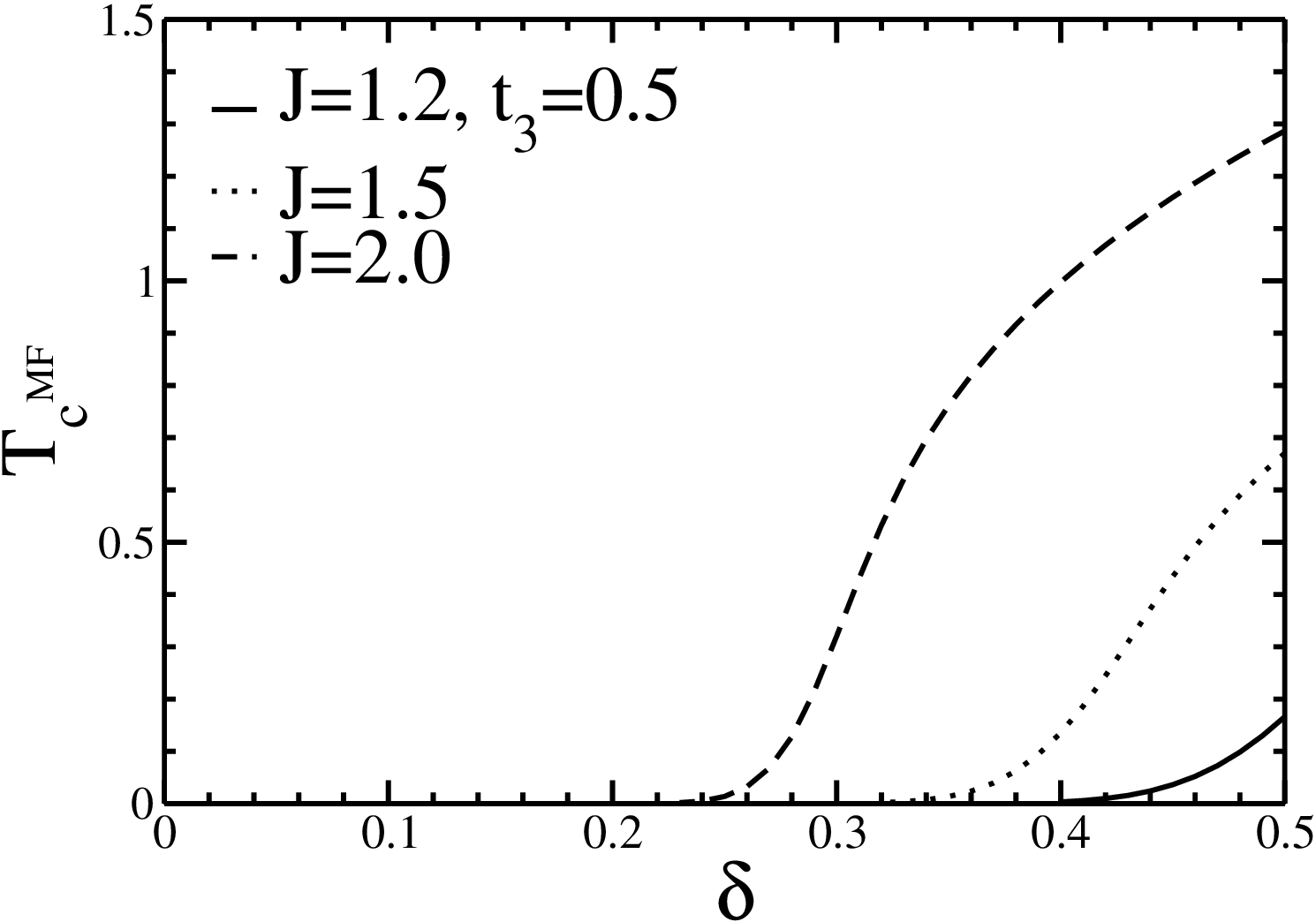}}
\caption{The mean-field critical temperature as a function of
carrier density at $t_{3}=0.5$.}
\label{fig:3}
\end{figure}

\begin{figure}[h]
\centering{
\includegraphics[width=6.0cm]{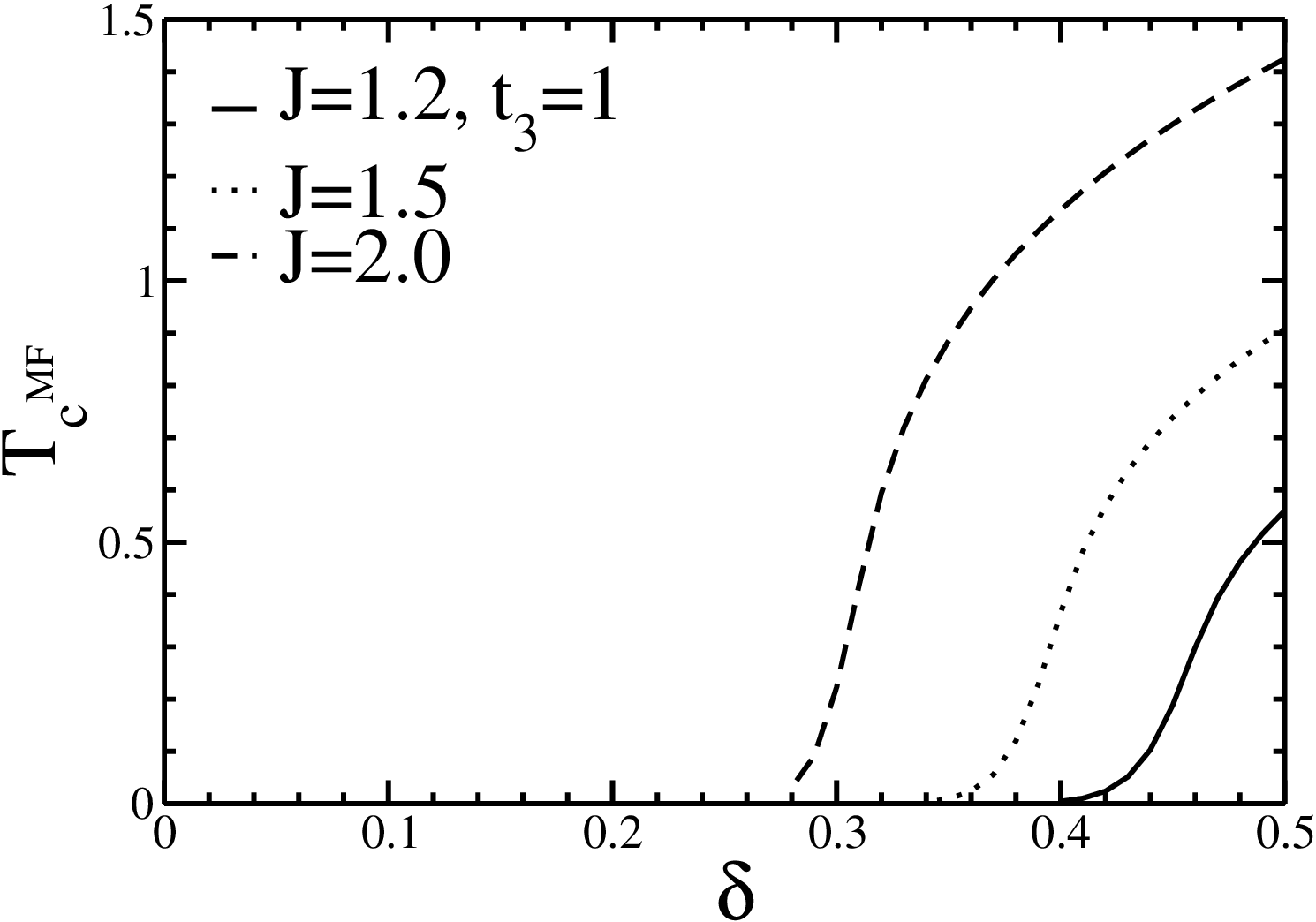}}
\caption{The same as in Fig.3 at $t_{3}=1$.}
\label{fig:4}
\end{figure}

It is important that superconductivity is suppressed at low carrier
densities in the $d$-wave pairing channel. To understand this
qualitatively, one can notice that at low doping, the function
$\gamma_{d}^{2}({\bf q})$ in the equation for the mean-field
critical temperature (\ref{gapequationT}) can be approximated as
$\gamma_{\bf q}\simeq q^{2}\cos(2\varphi_{{\bf q}})/2 \sim
m^{*}\epsilon_{F}\cos(2\varphi_{{\bf q}_{F}})/2$, where
$\epsilon_{F}$ is the Fermi energy and $m^{*}$ is an effective
quasiparticle mass, which is directly connected with $t_{2}$ and
$t_{3}$ (see the next Subsection). In this case,
Eq.~(\ref{gapequation2}) has the form of a standard BSC equation for
the critical temperature in the s-wave pairing channel with the
renormalized coupling $J\rightarrow Jm^{*2}\epsilon_{F}^{2}/4$.
Therefore, the mean field critical temperature is exponentially
small function of square of the carrier density $\delta^{2} \sim
\epsilon_{F}^{2}$ in this case.

\subsection{The critical temperature}

In order to study the BKT transition in the system, it is necessary
to consider the superconducting order parameter transfer phase
fluctuations. The phase of the order parameter can be introduced in
analogy with the fermion case, where the phase $\theta_{{\bf n}{\bf
m}}$ of the fermion operator can be presented as a product of the
neutral operator $\chi (\tau , {\bf n})$ and its phase $\exp
\{i\theta (\tau , {\bf n})/2\}$ \cite{PhysRep}. In our case:
$$
X_{{\bf n}}^{2,1/2}=(X_{{\bf n}}^{1/2,2})^{\dagger} =\chi (\tau ,
{\bf n})\exp \{i\theta (\tau , {\bf n})/2\}.
$$
In this case, the superconducting order parameter can be presented
as a product of its amplitude and the phase:
\begin{equation}
\langle X_{{\bf n}}^{2,1/2}X_{{\bf m}}^{2,1/2}
\rangle=\delta_{{\bf n},{\bf m}+{\bf \rho}}
\Delta_{{\bf n}{\bf m}}\exp (i\theta_{{\bf n}{\bf m}}) .
\label{gapmodulusphase}
\end{equation}

It can be shown that the phase-dependence of the thermodynamic
potential of the fermion system with the Green function (\ref{GF})
and the gap function (\ref{gapmodulusphase}) in the limit of small
fluctuations of the phase of the order parameter is equal to
\begin{equation}
\Omega (\Delta , \theta)= \frac{{\cal J}}{2}\int d^{2}r
({\bf\nabla}\theta )^{2}, \label{Omega}
\end{equation}
where the stiffness in the long-wave limit is
\begin{eqnarray}
{\cal J}=\frac{\delta}{4m^{*}} -\frac{1}{16m^{*2}}\frac{1}{T} \int
\frac{d^{2}k}{(2\pi )^{2}} \frac{{\bf
k}^{2}}{\cosh^{2}[\sqrt{\varepsilon ({\bf k})^{2}+\Delta_{d}^{2}
\gamma_{d}^{2}({\bf k})}/2T]} . \label{J}
\end{eqnarray}
(see, for example \cite{Loktev2}). In our case, the effective mass
of the free quasi-particles is $m^{*}=1/[4(t_{2}+2t_{3})]$.

In analogy with the 2D spin $XY$-model \cite{Izyumov},
the equation for the BKT transition
critical temperature, below which the phases of order parameter
(the spin $\nabla\theta$ orientation in the $XY$-model case) become
algebraically ordered has the following form:
\begin{equation}
T_{c}=\frac{\pi}{2}{\cal J}(\Delta_{d} ,\mu ', T_{c}), \label{TBKT}
\end{equation}
where function ${\cal J}$ is defined in (\ref{J}).

The doping dependence of the superconducting critical temperature
$T_{c}$ can be found by solving the system of
Eqs.~(\ref{gapequationT}), (\ref{numberequationT}) and (\ref{TBKT}).
The solution of this set at different values of interaction shows
that the doping-dependence of $T_{c}$ has qualitatively the same
form as the doping-dependence of $T_{c}^{MF}$ (Fig.4). It is
possible to study some limiting cases of the solution of
Eq.~(\ref{TBKT}) analytically. In particular, in the limit of rather
large carrier densities, when $\Delta_{d}\ll T_{c}$ ($T_{c}$ is
close to $T_{c}^{MF}$), one can make an expansion in powers of
$\Delta_{d}/T_{c}$ on the right hand side of Eq.~(\ref{TBKT}). In
this case, this equation transforms to:
\begin{equation}
T_{c}^{3}=A\Delta_{d}^{2}(T_{c})\delta^{3}/m^{*}, \label{TBKT2}
\end{equation}
where $A\simeq \pi^{3}/128$. Since the gap parameter depends on the
critical temperature as
$\Delta_{d}(T)=\Delta_{d}(0)[1-(T/T_{c}^{MF})^{2}]^{1/\alpha}$,
where $\alpha\geq 1$, at temperatures close to the mean-field
critical temperature \cite{IJMP}, the solution of Eq.~(\ref{TBKT2})
is:
\begin{equation}
T_{c}\simeq T_{c}^{MF}\left[ 1-\frac{1}{2}\left[
\frac{m^{*}T_{c}^{MF3}}{A\Delta_{d}^{2}(0)} \right]^{\alpha
/2}\frac{1}{\delta^{3\alpha /2}} \right] . \label{TBKTsol}
\end{equation}
In other words, the critical temperature approaches to the
mean-field critical temperature as the doping increases at large
carrier densities. It is interesting, that equation (\ref{TBKT2}) is
also valid in the limit of low carrier concentrations when both
critical temperatures are suppressed. Therefore, the second term in
Eq.~(\ref{J}) is important at any carrier concentration in the
d-wave pairing channel, contrary to the s-pairing case, where this
term can be omitted at low values of $\delta$ and $T_{c}\sim
\delta$.

It is important that the amplitude of $T_{c}$ is much smaller than
$T_{c}^{MF}$, and the pseudogap region $T_{c}<T<T_{c}^{MF}$ is
rather large in this case (Fig.4). This is similar to the phase
diagram of cuprates, where there is a large pseudogap region above
the critical temperature at low carrier densities. It must be
stressed, that in the case of higher carrier densities, one must
condsider a model with more complicated effective interaction,
comparing to Eq.~(\ref{Hattr}).

In order to get superconductivity to develop at $\delta\simeq 0.05$,
similar to some of HTSCs, one needs to choose a pretty large
interaction energy $J>5t_{2}$, which correspond to the energy scale
of $1eV$. However, to make a quantitative comparison with
experiments, one needs to take into account different effects, which
were omitted in this paper. In particular, optical phonons can give
a significant contribution into the electron-electron interaction in
cuprates, which can lead to decreasing of the required values of $J$
in order to get superconductivity at $\delta\simeq 0.05$.

\section{Conclusion}

To conclude, we have studied the superconducting properties of an
effective model introduced in \cite{Loktev} in order to describe low
carrier density properties of HTSCs. It was shown that the $d$-wave
pairing superconductivity in this model exists when the carrier
density is larger then some critical density. This critical density
strongly depends on the interaction energy, and it is growing with
interaction decreasing. The amplitude of the pseudogap temperature
is much larger than the critical temperature values, which resembles
the experimental situation on cuprates. There are some open issues
which must be resolved. First of all, it is important to understand
how to generalize the results on the case of larger carrier
densities. In this case, the antiferromagnetic sublattice breaks
down and it is not enough to put the attraction to be equal to the
antiferromagnetic bond energy $J$. The effective antiferromagnetic
attraction decreases. Also, the doping increasing is accompanied by
increasing of the number of scattering centers created by dopants,
which also leads to suppression of superconductivity (see, for
example \cite{IJMP}). Another important problem which is widely
discussed nowadays is to understand, whether the model can give an
inhomogeneous superconducting state (see, e.g. \cite{Kivelson} and
references therein). It is also necessary to estimate pairings in
other channels with different symmetry of the order parameter. These
and some other questions are planned to be studied in the nearest
future.



\begin{thebibliography}{99}

\bibitem{Damascelli}
A.~Damascelli, Z.~Hussain, and Z.-X.~Shen, Rev.~Mod.~Phys.~{\bf
75},473 (2003).

\bibitem{Franck}
J.P.~Franck,
in {\it Physical Properties of High Temperature Superconductors IV},
ed. D.M.~Ginsberg, World Scientific, Singapore, P.189 (1994).

\bibitem{Shen}
Z.-X.~Shen and D.S.~Dessau,
Phys.~Rep. {\bf 253}, 1 (1995).

\bibitem{Ding}
H.~Ding, T.~Yokota, J.C.~Campuzano et al,
Nature (London) {\bf
382},51 (1996).

\bibitem{Loeser}
A.G.~Loeser, Z.-X.~Shen, D.S.~Dessau, Science {\bf 273}, 325 (1996).

\bibitem{Lee}
P.A.~Lee, N.~Nagaosa, and X.-G.~Wen, Rev.~Mod.~Phys.~{\bf 78}, 17
(2006).

\bibitem{Anderson}
P.W.~Anderson, Science {\bf 235}, 1196 (1987).

\bibitem{Zhang}
F.C.~Zhang and T.M.~Rice, Phys. Rev. B {\bf 37}, 3759 (1988).

\bibitem{SchmittRink}
S. Schmitt-Rink, C. M. Varma, and A. E. Ruckenstein, Phys. Rev.
Lett. {\bf 60}, 2793 (1988).

\bibitem{Kane}
C. L. Kane, P. A. Lee, and N. Read, Phys. Rev. B {\bf 39}, 6880
(1989).

\bibitem{Martinez}
G. Martinez and P. Horsch, Phys. Rev. B {\bf 44}, 317 (1991).

\bibitem{Plakida}
N. M. Plakida and V. S. Oudovenko, Phys. Rev. B {\bf 59}, 11949
(1999).

\bibitem{Bensimon}
D.~Bensimon, R.~Zeyher, preprint cond-mat/0603585.

\bibitem{Haule}
K.~Haule, G.~Kotliar, preprint cond-mat/0605149.

\bibitem{Emery}
V.J.~Emery, Phys.~Rev.~Lett.~{\bf 58}, 2794 (1987).

\bibitem{Loktev}
V.M.~Loktev,
Fiz.~Nizk.~Temp. {\bf 31}, 1 (2005).

\bibitem{Babichenko}
V.S.~Babichenko, Yu.~Kagan,
JETP~Lett. {\bf 56}, 305 (1992).

\bibitem{Trugman}
S.A.~Trugman, Phys.~Rev.~B {\bf 37}, 1597 (1988).

\bibitem{Scalapino}
D.J.~Scalapino and S.A.~Trugman,
Philos.~Mag.~B {\bf 74}, 607(1996).

\bibitem{Mancini}
F.~Mancini, A.~Avella, Adv.~Phys. {\bf 53}, 537 (2004).

\bibitem{PhysRep}
V.M.~Loktev, R.M.~Quick, S.G.~Sharapov, Phys.~Rep. {\bf 349}, 2
(2001).

\bibitem{Loktev2}
S.G.~Sharapov, H.~Beck, V.M.~Loktev,
Phys.~Rev.~B {\bf 64}, 134519 (2001).

\bibitem{Izyumov}
Yu.A.~Izyumov, Yu.M.~Skryabin,
{\it Statistical Mechanics of Magnetically Ordered Systems},
Plenum, New York (1988).

\bibitem{IJMP}
V.M.~Loktev, V.~Turkowski,
Int.~Journ.~of~Mod.~Phys. B {\bf 18}, 2035 (2004).

\bibitem{PhysicaC}
V.M.Loktev and V.Turkowski,
Physica C {\bf 383}, 256 (2002).

\bibitem{Kivelson}
W.-F.~Tsai and S.~Kivelson,
Phys.~Rev.~B {\bf 73}, 214510 (2006).

\end{thebibliography}
\end{document}